# LANTERN: learn analysis transform network for dynamic magnetic resonance imaging with small dataset

Shanshan Wang, Yanxia Chen, Taohui Xiao, Ziwen Ke, Qiegen Liu, Hairong Zheng*

# Abstract

This paper proposes to learn analysis transform network for dynamic magnetic resonance imaging (LANTERN) with small dataset. Integrating the strength of CS-MRI and deep learning, the proposed framework is highlighted in three components: (i) The spatial and temporal domains are sparsely constrained by using adaptively trained CNN. (ii) We introduce an end-to-end framework to learn the parameters in LANTERN to solve the difficulty of parameter selection in traditional methods. (iii) Compared to existing deep learning reconstruction methods, our reconstruction accuracy is better when the amount of data is limited. Our model is able to fully exploit the redundancy in spatial and temporal of dynamic MR images. We performed quantitative and qualitative analysis of cardiac datasets at different acceleration factors (2x-11x) and different undersampling modes. In comparison with state-of-the-art methods, extensive experiments show that our method achieves consistent better reconstruction performance on the MRI reconstruction in terms of three quantitative metrics (PSNR, SSIM and HFEN) under different undersamling patterns and acceleration factors.

# Introduction

Dynamic magnetic resonance imaging is able to provide important anatomical and functional information in a spatial-temporal manner. However, a fundamental challenge of MRI is its slow imaging speed a.k.a long imaging time, which hinders its wide applications. To address this challenge, there have been different efforts devoted by researchers ranging from prompting hardware to software developments such as parallel imaging using phased array coils [1], fast imaging sequences [2], and reduced-scan techniques with advanced image reconstruction algorithms.

Our specific focus here is the signal processing based MR image reconstruction from incomplete k-space data. Since the acquisition time of k-space is proportional to its amount, this strategy accelerates MR scan by undersampling or partial sampling of k-space. Undersampling introduces violations of the Nyquist sampling theorem and may cause aliasing and blurring issues by direct inverse Fourier transform [3], [4]. To solve these issues, prior knowledges are normally incorporated in the reconstruction formulation as regulations. Specifically, under the support of the well-known compressed sensing (CS) theory, researchers have developed a series of dynamic image reconstruction methods by exploiting either spatial or temporal redundancy or both with different sampling patterns. For example, with a random k-t sampling pattern, k-t FOCUSS [5] was proposed, whose special cases included the celebrated k-t BLAST and k-t SENSE [6]. There was also a k-t iterative support detection (k-t ISD) method to improve the CS dynamic MR imaging methods [7]. These methods along others have explored different sparsifying transforms such as total variation (TV) and wavelet in spatial domain [8]–[10]; Fourier transform [11], [12], finite difference [13], and principal component analysis [14], [15] in the temporal domain; and 3D transforms such as wavelet-Fourier transform [16] or 3D wavelet transform in the spatial-temporal domain [17]. Besides the predefined transforms, dictionary learning has also been investigated. For example, temporal gradient sparsity was explored by [3] with adaptively trained dictionary and a patch-based 3-D spatiotemporal dictionary was trained for sparse representations of the dynamic image sequence [18]. In addition to the sparsifying transforms, low rank has been utilized to complete missing or corrupted entries for a matrix as well. Typical instances include Bo Zhao et al proposed combine PS and sparsity constraints to improve MR reconstruction performance[19] and the L+S [20], k-t SLR [13] methods. These methods all made great contributions to dynamic MR imaging. Nevertheless, the prior knowledges utilized are still limited to few samples or reference images [21]. Furthermore, the iterative reconstruction can be time-consuming with parameters hard to tune.

Deep learning based MR image reconstruction is an emerging field to accelerate MR scan. There are model-based deep learning methods that formulate the prior regularization iterative reconstruction process into network learning process. Typical examples include variational network (VN-net) [22], alternating direction method of multipliers network (ADMM-net) [23] and Model DL [24], etc. [21]. In addition to the model based methods, there are also direct end-to-end learning techniques that identify the mapping relationship between the undersampled and fully-sampled pairs. Instances consist of AUTOMAP [25], U-net [26], KIKI-net [27], recursive dilated net [28], and so on so forth [29]. Among all of them, there are only a few end-to-end learning networks for dynamic MR imaging [4], [30]. These works directly learn the mapping relationship and have shown great experimental results. Nevertheless, the explanation of this work is more empirical, which hasn't taken advantage of the theoretically explainable compressed sensing framework.

To bridge the gap between the models based dynamic imaging work and the empirical direct map learning framework, this work proposes a convolutional analysis transform network learning for dynamic magnetic resonance imaging with convolutional dubbed as LANTERN. Specifically, we initialize the discrete cosine transform（DCT）in the spatial domain and total variation (TV) in the temporal domain to fully exploit the redundancy of dynamic image sequnces. To optimize the models, we use Alternating Direction Method of Multipliers (ADMM) which is a valid variable separable method. Nevertheless, the reconstruction is normally time-consuming and its parameters have to be hand-tuned. Inspired by the convolutional neural network, we use CNN to learn all the parameters in the above formula, and then use the trained model for previously unseen data. We only collect part of the k-space data for reconstruction this can reduce the acquisition time and our model is trained offline, so it only requires a short reconstruction time. Our experiments demonstrate that the proposed scheme can effective reconstruct dynamic MR image with high accuracy and fast speed. Compared with the state-of-the-art method, D5C5 [4], k-t SLR [13], our method presents superior performance in both quantitative and qualitative analysis.

# Methods

## Dynamic imaging model

In dynamic MR imaging, the measured signal $\mathbf{y}_t \in \mathbb{C}^M$ at time t can be described as follows

$$\mathbf{y}_t = \boldsymbol{F}_t \boldsymbol{x}_t + \boldsymbol{\eta}_t \tag{1}$$

where $\boldsymbol{F}_t \in \mathbb{C}^{M*N}$ is the measurement matrix and $\boldsymbol{\eta}_t \in \mathbb{C}^M$ is the measurement noise for the t-th vectorized cardiac phase image $\boldsymbol{x}_t \in \mathbb{C}^N$; $\boldsymbol{F}_t = \boldsymbol{P}_t \boldsymbol{F}_{2D}$; $\boldsymbol{P}_t$ is an $M \times N$ undersampling matrix whose rows are extracted from an $N \times N$ identity matrix according to the k-space sampling locations at time t ($M \ll N$). $\boldsymbol{F}_{2D} \in \mathbb{C}^{N \times N}$ is the unitary matrix representing the 2D Fourier transform. Suppose a total of Q cardiac phases are acquired, the entire acquisition process can be described as follows:

$$\boldsymbol{y} = \mathbf{F_u} \boldsymbol{x} + \boldsymbol{e} \tag{2}$$

Where $\boldsymbol{x} = [\boldsymbol{x}_1^H, \boldsymbol{x}_2^H, \ldots \boldsymbol{x}_t^H \ldots \boldsymbol{x}_Q^H] \in \mathbb{C}^{NQ*1}$ represents the stacked Q phase images; $\boldsymbol{e} = [\boldsymbol{\eta}_1^H, \boldsymbol{\eta}_2^H, \ldots \boldsymbol{\eta}_t^H \ldots \boldsymbol{\eta}_Q^H] \in \mathbb{C}^{MQ}$; $\boldsymbol{y} = [\boldsymbol{y}_1^H, \boldsymbol{y}_2^H, \ldots \boldsymbol{y}_t^H \ldots \boldsymbol{y}_Q^H] \in \mathbb{C}^{MQ}$ is the under-sampled k-space data, and $\mathbf{F_u} = diag\{\boldsymbol{F}_1, \boldsymbol{F}_2, \ldots \boldsymbol{F}_t, \ldots, \boldsymbol{F}_Q\} \in \mathbb{C}^{MQ*NQ}$. H is the hermitan transpose operation.

## The proposed method

1) **Sparse convolutional coding feature preserving prior model**
   The recovery of $\boldsymbol{x}$ from $\boldsymbol{y}$ is an underdetermined problem because $M \ll N$. To reconstruct $\boldsymbol{x}$, we introduce a sparse coding convolutional feature preserving model to overcome the ill-posedness nature and propose the following model

$$\arg\min_{\boldsymbol{x}} \left\{ \frac{1}{2} \|\mathbf{F_u}\boldsymbol{x} - \boldsymbol{y}\|_2^2 + \sum_{l=1}^{L} \lambda_l Pri(\boldsymbol{\Phi}_l \boldsymbol{x}) \right\} \tag{3}$$

where $\frac{1}{2}\|\mathbf{F_u}\boldsymbol{x} - \boldsymbol{y}\|_2^2$ is the data fidelity term, $Pri(\cdot)$ denotes prior regularization function derived from data with sparse coding convolutional operator $\boldsymbol{\Phi}_l$ extracting image features and $\lambda_l$ means regularization parameter. $L$ represents the number of filters. To solve this, we introduce auxiliary variables $\mathbf{v} = [\mathbf{v}_1^T, \mathbf{v}_2^T, \ldots \mathbf{v}_t^T \ldots, \mathbf{v}_Q^T]^T \in \mathbb{C}^{NQ \times 1}$ and get the following constrained formulation

$$\arg\min_{\boldsymbol{x},\mathbf{v}} \frac{1}{2} \|\mathbf{F_u}\boldsymbol{x} - \boldsymbol{y}\|_2^2 + \sum_{l=1}^{L} \lambda_l Pri(\boldsymbol{\Phi}_l \mathbf{x}) \quad \text{s.t.} \quad \mathbf{v} = \boldsymbol{x} \tag{4}$$

Adopting augmented Lagrangian technique, the constrained problem in (4) can be transformed into the following unconstrained one:

$$\mathcal{L}_\rho(\boldsymbol{x},\mathbf{v},\boldsymbol{\alpha}) = \frac{1}{2}\|\mathbf{F_u}\boldsymbol{x} - \boldsymbol{y}\|_2^2 + \sum_{l=1}^{L} \lambda_l Pri(\boldsymbol{\Phi}_l \boldsymbol{x}) - \langle \boldsymbol{\alpha}, \mathbf{v} - \boldsymbol{x} \rangle + \frac{\rho}{2} \|\mathbf{v} - \boldsymbol{x}\|_2^2 \tag{5}$$

Where $\alpha$ are Lagrangian multipliers; $\rho$ represents the scaling factor; the above optimization problem can be further divided into three subproblems by using the alternating direction multiplier method with an assistant variable $\beta$ introduced:

$$\begin{cases} x^{i+1} = \underset{x}{\mathrm{argmin}}\, \frac{1}{2}\|F_u x - y\|_2^2 + \frac{\rho}{2}\|x + \beta^i - v^i\|_2^2 \\ v^{i+1} = \underset{v}{\mathrm{arg\,min}}\, \frac{\rho}{2}\|x^{i+1} + \beta^i - v\|_2^2 + \sum_{l=1}^L \lambda_l Pri(\Phi_l v) \\ \beta^{i+1} = \beta^i + \tilde{\eta}(x^{i+1} - v^{i+1}) \end{cases} \quad (6)$$

**2) Alternating direction minimization algorithm**

**(a) Subproblem $x$.** Adopting least squares to solve the first equation in Eq. (6), we have

$$(F_u^H F_u + \rho I)x = F_u^H y + \rho(v - \beta)$$

Then further let $F_u = PF$, with $P = \mathrm{diag}\{P_1, P_2, \ldots P_t, \ldots P_Q\} \in R^{MQ \times NQ}$ and $F = \mathrm{diag}\{F_{2D}, F_{2D}, \ldots, F_{2D}\} \in \mathbb{C}^{NQ \times NQ}$, we have the following solution

$$x = F^H(P^H P + \rho I)^{-1}[P^H y + \rho F(v - \beta)]$$

Where $P^H P$ is a diagonal matrix that can be quickly calculated.

**(b) Subproblem $v$.** For the update of $v$, we adopts the gradient descent method. With the gradient $\nabla v = \rho(v - x - \beta) + \sum_{l=1}^L \lambda_l \Phi_l^H \mathcal{H}_{pri}(\Phi_l v)$, we have

$$v^{(i+1)} = v^{(i)} - l_r \nabla v = (1 - l_r \rho) v^{(i)} + l_r \rho(x + \beta) - \sum_{l=1}^L \lambda_l l_r \Phi_l^H \mathcal{H}_{pri}(\Phi_l v)$$

Where $\mathcal{H}_{pri}$ means the derivative of the prior regulation function $Pri$ and $l_r$ is the step size. Therefore, we have

$$\begin{cases} x^{(i)} = F^H(P^H P + \rho^{(n)} I)^{-1}[P^H y + \rho^{(n)} F(v^{(n-1)} - \beta^{(n-1)})] \\ v^{(n,k)} = \mu_1^{(n,k)} v^{(n-1)} + \mu_2^{(n,k)}(x^{(n)} + \beta^{(n-1)}) - \sum_{l=1}^L \tilde{\lambda}_l \Phi_l^H \mathcal{H}_{pri}(\Phi_l v^{(n-1)}) \\ \beta^{(n)} = \beta^{(n-1)} + \tilde{\eta}^{(n)}(x^{(n)} - v^{(n)}) \end{cases} \quad (7)$$

Where, $u_1 = 1 - l_r \rho, u_2 = l_r \rho, \tilde{\lambda}_l = \lambda_l l_r$.

Rewrite the above formula (7) to the following formula (8), split $v^{(n,k)}$ into the $Addition$ layer, $Conv1$ layer, $Nonlinear$ layer and $Conv2$ layers. In particular, we consider filter $\Phi$ as convolution kernel. And $\mathcal{H}_{pri}$ is approximated by learning a piecewise linear function $S_{PLF}(\cdot)$.

$$\begin{cases} Recon:\ x^{(n)} = F^H(P^H P + \rho^{(n)} I)^{-1}[P^H y + \rho^{(n)} F(v^{(n-1)} - \beta^{(n-1)})] \\ Addition:\ v^{(n,k)} = \mu_1^{(n,k)} v^{(n,k-1)} + \mu_2^{(n,k)}(x^{(n)} + \beta^{(n-1)}) - C_2^{(n,k)} \\ Conv1:\ C_1^{(n,k)} = \sum_{l=1}^L (w_{1,l}^{(n,k)} * v^{(n,k-1)} + b_{1,l}^{(n,k)}) \\ Nonlinear:\ h^{(n,k)} = S_{PLF}(C_1^{(n,k)}; \{p_i, q_i^{(n,k)}\}_{i=1}^{N_c}) \\ Conv2:\ C_2^{(n,k)} = \sum_{l=1}^L (w_{2,l}^{(n,k)} * h^{(n,k)} + b_{2,l}^{(n,k)}) \\ Multi:\ \beta^{(n)} = \beta^{(n-1)} + \tilde{\eta}^{(n)}(x^{(n)} - v^{(n)}) \end{cases} \quad (8)$$

## LANTERN network architecture

In order to exploit the extensive temporal and spatial redundancy of dynamic magnetic resonance imaging, we propose a LANTERN architecture was loosely inspired by [23], which adaptively learns sparse convolution kernels and regularization parameters through convolutional neural networks without artificial adjustment. The flow of the network is shown in Fig.1 (A). The input is an undersampled k-space data. After N iterations, a reconstructed image can be obtained. The reconstructed image and label are used to calculate the mean square error, and then the back-propagation is used to update the parameters in the network. This way, you can get a high-quality, artifact-free image that is close to Ground Truth. Fig.1 (B) shows the specific process of the nth iteration, each layer corresponding to the following forward propagation formula, wherein the detail process of the $prior$ layer is described in detail in Fig.1 (c). The $prior$ layer includes $Conv1$, $Nonlinear$ and $Conv2$, where $Conv1$ and $Conv2$ learn filter operators that make the image sparse. A piecewise linear function is used to approximate the derivative of the regularization function.

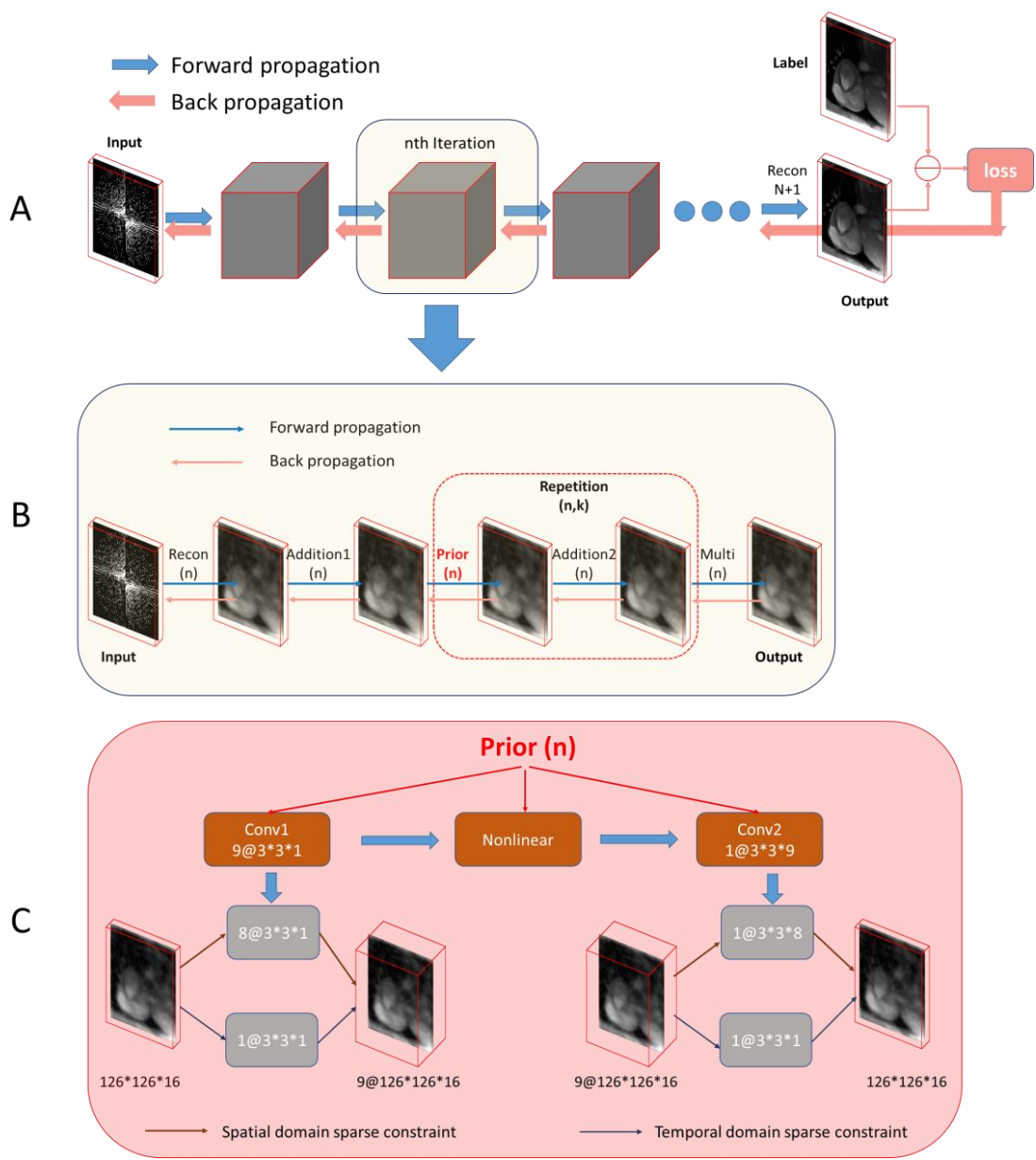

Fig. 1. The proposed LANTERN network architecture for dMRI reconstruction. In (A) and (B), the blue arrow indicates the process of reconstructing the undersampled image by forward propagation. The pink arrow indicates the process of back-propagation the update parameters.

Where, the n represents the nth iteration and the N represents a total of N iterations. The K express the priori loop for k times and (n, k) means that in the nth iteration, the a priori loops k times.

### 1) The formula for the propagation of each layer is as follows

In Fig.1 (b) and (c), the $Recon$ stands for reconstruction layer $x^{(n)}$. The input is undersampled k-space data $y$ and $\mathbf{v}^{(n-1)}, \mathbf{\beta}^{(n-1)}$. $prior$ Stands for addition layer $\mathbf{v}^{(n,k)}$. Performs simple weighted summation operation. The input is $\mathbf{v}^{(n,k-1)}, x^{(n)}, \mathbf{\beta}^{(n-1)}$ and $\mathbf{C}_2^{(n,k)}$. $Conv1$ And $Conv2$ stands for convolution layer. The input is $\mathbf{v}^{(n,k-1)}$ and $\mathbf{h}^{(n,k)}$. In which we used 3D convolution in the spatial domain, mainly for feature extraction, however, the parameter amount of 3D convolution is relatively large. In order to reduce the parameter quantity, we use 2D convolution in the time domain TV filter. $Nonlinear$ Stands for nonlinear layer $\mathbf{h}^{(n,k)}$. The input is $\mathbf{C}_1^{(n,k)}, \{p_i, q_i^{(n,k)}\}_{i=1}^{N_c}$. $Multi$ Stands for multiplier update layer $\mathbf{\beta}^{(n)}$. The input is $\mathbf{\beta}^{(n-1)}$, $x^{(n)}$ and $\mathbf{v}^{(n)}$. The loss layer can be calculated from the reconstructed dMRI and the original fully sampled dMRI by a standard mean square error. Note that the Recon layer includes three different forward propagation equations $R_{org}$, $R_{mid}$ and $R_{final}$, due to when n=1 the input $\mathbf{v}^{(n-1)}, \mathbf{\beta}^{(n-1)}$ is 0. So, there are different iteration formulas as n increases, but they are all derived from formula (8). Similarly, $Addition$ and $Multi$ layers are also different in formula because of the number of iterations of the input, which can be derived from formula (8).

$$Recon: \quad R_{org}: x^{(n)} = F^H \left(P^H P + \rho^{(n)} I\right)^{-1} (P^H y)$$

$$R_{mid}\ and\ R_{final}: x^{(n)} = F^H \left(P^H P + \rho^{(n)} I\right)^{-1} \left[P^H y + \rho^{(n)} F\left(\mathbf{v}^{(n-1)} - \mathbf{\beta}^{(n-1)}\right)\right]$$

$$Addition: \quad A_{org}: \mathbf{v}^{(1,0)} = \mu_2^{(1,0)} x^{(1)}, \text{ initialization } \mathbf{C}_2^{(1,0)} = 0$$

$$A_{mid}: \mathbf{v}^{(1,k)} = \mu_1^{(1,k)} \mathbf{v}^{(1,k-1)} + \mu_2^{(1,k)} x^{(1)} - \mathbf{C}_2^{(1,k)}$$

$$A1_{org}: \mathbf{v}^{(n,0)} = \mu_2^{(n,0)} \left(x^{(n)} + \mathbf{\beta}^{(n-1)}\right)$$

$$A1_{mid}: \mathbf{v}^{(n,k)} = \mu_1^{(n,k)} \mathbf{v}^{(n,k-1)} + \mu_2^{(n,k)} \left(x^{(n)} + \mathbf{\beta}^{(n-1)}\right) - \mathbf{C}_2^{(n,k)}$$

$$Conv1: \quad \mathbf{C}_1^{(n,k)} = \sum_{l=1}^{L} \left(\mathbf{w}_{1,l}^{(n,k)} * \mathbf{v}^{(n,k-1)} + \mathbf{b}_{1,l}^{(n,k)}\right)$$

$$Nonlinear: \quad \mathbf{h}^{(n,k)} = S_{PLF}\left(\mathbf{C}_1^{(n,k)}; \{p_i, q_i^{(n,k)}\}_{i=1}^{N_c}\right)$$

$$Conv2: \quad \mathbf{C}_2^{(n,k)} = \sum_{l=1}^{L} \left(\mathbf{w}_{2,l}^{(n,k)} * \mathbf{h}^{(n,k)} + \mathbf{b}_{2,l}^{(n,k)}\right)$$

$$Multi: \quad M_{org}: \mathbf{\beta}^{(1)} = \tilde{\eta}^{(1)} \left(x^{(1)} - \mathbf{v}^{(1)}\right)$$

$$M_{mid}\ and\ M_{final}: \mathbf{\beta}^{(n)} = \mathbf{\beta}^{(n-1)} + \tilde{\eta}^{(n)} \left(x^{(n)} - \mathbf{v}^{(n)}\right)$$

Loss: $E(\Theta) = \frac{1}{|\Lambda|} \sum_{(y,x^{gt}) \in \Lambda} \frac{\sqrt{\|x(y,\Theta) - x^{gt}\|_2^2}}{\sqrt{\|x^{gt}\|_2^2}}$

Where, $\Lambda$ represents the number of training sets. $x(y,\Theta)$ is the LANTERN network output based on under-sampled data y in k-space and parameter $\Theta$ and $x^{gt}$ is fully sampled dMRI.

2) **The process of back-propagation parameter update is as follows**

As shown in Fig.1. In the process of back-propagation, the gradient is calculated in an opposite order. So, the gradient of the loss layer is as follows:

$$\frac{\partial E}{\partial x^{(n)T}} = \frac{1}{|\Lambda|} \frac{(x^{(n)} - x^{gt})}{\sqrt{\|x^{gt}\|_2^2}\sqrt{\|x^{(n)} - x^{gt}\|_2^2}}$$

(a) The gradient of the $Recon\ layer$ is as follows, update parameter $\rho^{(n)}$.
   N represents a total of N iterations.

$$\frac{\partial E}{\partial \rho^{(n)}} = \frac{\partial E}{\partial x^{(n)H}} \frac{\partial x^{(n)}}{\partial \rho^{(n)}}$$

Where,

$$\frac{\partial E}{\partial x^{(n)H}} = \begin{cases} \frac{\partial E}{\partial \beta^{(n)H}} \frac{\partial \beta^{(n)}}{\partial x^{(n)}} + \frac{\partial E}{\partial v^{(n,k)H}} \frac{\partial v^{(n,k)}}{\partial x^{(n)}} & n < N \\ \frac{1}{|\Lambda|} \frac{(x^{(n)} - x^{gt})}{\sqrt{\|x^{gt}\|_2^2}\sqrt{\|x^{(n)} - x^{gt}\|_2^2}} & n = N \end{cases}$$

$$\frac{\partial x^{(n)}}{\partial \rho^{(n)}} = F^H\{\left(-\frac{1}{[(P^H P + \rho^{(n)} I)^{-1}]^2}\right)[P^H y + \rho^{(n)} F(v^{(n-1)} - \beta^{(n-1)})]$$

$$+ (P^H P + \rho^{(n)} I)^{-1} F(v^{(n-1)} - \beta^{(n-1)})\}$$

When n=1, there is $\frac{\partial x^{(1)}}{\partial \rho^{(1)}} = F^H\{\left(-\frac{1}{[(P^H P + \rho^{(n)} I)^{-1}]^2}\right)[P^H y]\}$

We also compute the gradient to its inputs:

$$\frac{\partial E}{\partial v^{(n-1)}} : \frac{\partial E}{\partial x^{(n)H}} \frac{\partial x^{(n)}}{\partial v^{(n-1)}} = F^H (P^H P + \rho^{(n)} I)^{-1} \rho^{(n)} F * \frac{\partial E}{\partial x^{(n)H}}$$

$$\frac{\partial E}{\partial \beta^{(n-1)}} : \frac{\partial E}{\partial x^{(n)H}} \frac{\partial x^{(n)}}{\partial \beta^{(n-1)}} = -F^H (P^H P + \rho^{(n)} I)^{-1} \rho^{(n)} F \frac{\partial E}{\partial x^{(n)H}}$$

(b) The gradient of the $Multi$ layer is as follows, update parameter $\tilde{\eta}^{(n)}$.

$$\frac{\partial E}{\partial \tilde{\eta}^{(n)}} = \frac{\partial E}{\partial \beta^{(n)H}} \frac{\partial B^{(n)}}{\partial \tilde{\eta}^{(n)}} = \frac{\partial E}{\partial \beta^{(n)H}} * (x^{(n)} - v^{(n)})$$

Where,

$$\frac{\partial E}{\partial \beta^{(n)H}} = \frac{\partial E}{\partial \beta^{(n+1)H}} \frac{\partial \beta^{(n+1)}}{\partial \beta^{(n)}} + \frac{\partial E}{\partial x^{(n+1)H}} \frac{\partial x^{(n+1)}}{\partial \beta^{(n)}} + \frac{\partial E}{\partial v^{(n+1,k)H}} \frac{\partial v^{(n+1,k)}}{\partial \beta^{(n)}}$$

We also compute the gradient to its inputs:

$$\frac{\partial E}{\partial \boldsymbol{\beta}^{(n-1)}} : \frac{\partial E}{\partial \boldsymbol{\beta}^{(n)H}} \frac{\partial \boldsymbol{\beta}^{(n)}}{\partial \boldsymbol{\beta}^{(n-1)}} = \frac{\partial E}{\partial \boldsymbol{\beta}^{(n)H}}$$

$$\frac{\partial E}{\partial x^{(n)}} : \frac{\partial E}{\partial \boldsymbol{\beta}^{(n)H}} \frac{\partial \boldsymbol{\beta}^{(n)}}{\partial x^{(n)}} = \frac{\partial E}{\partial \boldsymbol{\beta}^{(n)H}} * \tilde{\eta}^{(n)}$$

$$\frac{\partial E}{\partial \mathbf{v}^{(n)}} : \frac{\partial E}{\partial \boldsymbol{\beta}^{(n)H}} \frac{\partial \boldsymbol{\beta}^{(n)}}{\partial \mathbf{v}^{(n)}} = -\frac{\partial E}{\partial \boldsymbol{\beta}^{(n)H}} * \tilde{\eta}^{(n)}$$

(c) The gradient of the $Addition$ layer is as follows, update parameters $\mu_1^{(n,k)}, \mu_2^{(n,k)}$.

$$\frac{\partial E}{\partial \mu_1^{(n,k)}} : \frac{\partial E}{\partial \mathbf{v}^{(n,k)H}} \frac{\partial \mathbf{v}^{(n,k)}}{\partial \mu_1^{(n,k)}} = \frac{\partial E}{\partial \mathbf{v}^{(n,k)H}} * \mathbf{v}^{(n,k-1)}$$

$$\frac{\partial E}{\partial \mu_2^{(n,k)}} : \frac{\partial E}{\partial \mathbf{v}^{(n,k)H}} \frac{\partial \mathbf{v}^{(n,k)}}{\partial \mu_2^{(n,k)}} = \frac{\partial E}{\partial \mathbf{v}^{(n,k)H}} * \left(x^{(n)} + \boldsymbol{\beta}^{(n-1)}\right)$$

Where,

$$\frac{\partial E}{\partial \mathbf{v}^{(n,k)H}} = \frac{\partial E}{\partial x^{(n+1)H}} \frac{\partial x^{(n+1)}}{\partial \mathbf{v}^{(n)}} + \frac{\partial E}{\partial \mathbf{v}^{(n,k+1)H}} \frac{\partial \mathbf{v}^{(n,k+1)}}{\partial \mathbf{v}^{(n,k)}} + \frac{\partial E}{\partial C_1^{(n,k+1)H}} \frac{\partial C_1^{(n,k+1)}}{\partial \mathbf{v}^{(n,k)}} + \frac{\partial E}{\partial \boldsymbol{\beta}^{(n)H}} \frac{\partial \boldsymbol{\beta}^{(n)}}{\partial \mathbf{v}^{(n)}}$$

We also compute the gradient to its inputs:

$$\frac{\partial E}{\partial \mathbf{v}^{(n,k-1)}} : \frac{\partial E}{\partial \mathbf{v}^{(n,k)H}} \frac{\partial \mathbf{v}^{(n,k)}}{\partial \mathbf{v}^{(n,k-1)}} = \frac{\partial E}{\partial \mathbf{v}^{(n,k)H}} * \mu_1$$

$$\frac{\partial E}{\partial x^{(n)}} : \frac{\partial E}{\partial \mathbf{v}^{(n,k)H}} \frac{\partial \mathbf{v}^{(n,k)}}{\partial x^{(n)}} = \frac{\partial E}{\partial \mathbf{v}^{(n,k)H}} * \mu_2$$

$$\frac{\partial E}{\partial \boldsymbol{\beta}^{(n-1)}} : \frac{\partial E}{\partial \mathbf{v}^{(n,k)H}} \frac{\partial \mathbf{v}^{(n,k)}}{\partial \boldsymbol{\beta}^{(n-1)}} = \frac{\partial E}{\partial \mathbf{v}^{(n,k)H}} * \mu_2$$

$$\frac{\partial E}{\partial C_2^{(n,k)}} : \frac{\partial E}{\partial \mathbf{v}^{(n,k)H}} \frac{\partial \mathbf{v}^{(n,k)}}{\partial C_2^{(n,k)}} = -\frac{\partial E}{\partial \mathbf{v}^{(n,k)H}}$$

(d) The gradient of the $Conv2$ and $Conv1$ layer is as follows, update parameters $w_2^{(n,k)}, b_2^{(n,k)}$ and $w_1^{(n,k)}, b_1^{(n,k)}$.

$Conv2$ :

$$\frac{\partial E}{\partial \boldsymbol{w}_2^{(n,k)}} = \frac{\partial E}{\partial \boldsymbol{C}_2^{(n,k)H}} \frac{\partial \boldsymbol{C}_2^{(n,k)}}{\partial \boldsymbol{w}_2^{(n,k)}} \qquad \frac{\partial E}{\partial \boldsymbol{b}_2^{(n,k)}} : \frac{\partial E}{\partial \boldsymbol{C}_2^{(n,k)H}} \frac{\partial \boldsymbol{C}_2^{(n,k)}}{\partial \boldsymbol{b}_2^{(n,k)}}$$

Where,

$$\frac{\partial E}{\partial \boldsymbol{C}_2^{(n,k)H}} = \frac{\partial E}{\partial \boldsymbol{v}^{(n,k)H}} \frac{\partial \boldsymbol{v}^{(n,k)}}{\partial \boldsymbol{C}_2^{(n,k)H}}$$

We also compute the gradient to its inputs:

$$\frac{\partial E}{\partial \boldsymbol{h}^{(n,k)}} : \frac{\partial E}{\partial \boldsymbol{C}_2^{(n,k)H}} \frac{\partial \boldsymbol{C}_2^{(n,k)}}{\partial \boldsymbol{h}^{(n,k)}}$$

$Conv1$:

$$\frac{\partial E}{\partial \boldsymbol{w}_1^{(n,k)}} : \frac{\partial E}{\partial \boldsymbol{C}_1^{(n,k)H}} \frac{\partial \boldsymbol{C}_1^{(n,k)}}{\partial \boldsymbol{w}_1^{(n,k)}} \qquad \frac{\partial E}{\partial \boldsymbol{b}_1^{(n,k)}} : \frac{\partial E}{\partial \boldsymbol{C}_1^{(n,k)H}} \frac{\partial \boldsymbol{C}_1^{(n,k)}}{\partial \boldsymbol{b}_1^{(n,k)}}$$

Where,

$$\frac{\partial E}{\partial \boldsymbol{C}_1^{(n,k)H}} = \frac{\partial E}{\partial \boldsymbol{h}^{(n,k)H}} \frac{\partial \boldsymbol{h}^{(n,k)}}{\partial \boldsymbol{C}_1^{(n,k)T}}$$

We also compute the gradient to its inputs:

$$\frac{\partial E}{\partial \boldsymbol{v}^{(n,k-1)}} : \frac{\partial E}{\partial \boldsymbol{C}_1^{(n,k)H}} \frac{\partial \boldsymbol{C}_1^{(n,k)}}{\partial \boldsymbol{v}^{(n,k-1)}}$$

(e) **The gradient of the** $Nonlinear\ layer$ **is as follows, update parameter** $q_i^{(n,k)}$.

$$\frac{\partial E}{\partial \boldsymbol{q}_i^{(n,k)}} = \frac{\partial E}{\partial \boldsymbol{h}^{(n,k)H}} \frac{\partial \boldsymbol{h}^{(n,k)}}{\partial \boldsymbol{q}_i^{(n,k)}}$$

We also compute the gradient to its inputs:

$$\frac{\partial E}{\partial \boldsymbol{C}_1^{(n,k)}} : \frac{\partial E}{\partial \boldsymbol{h}^{(n,k)H}} \frac{\partial \boldsymbol{h}^{(n,k)}}{\partial \boldsymbol{C}_1^{(n,k)}}$$

# Experiments

## Parameter Setting of Network training

We input the undersampled k-space data and its corresponding fully-sampled dMRI into the network. We trained separate LANTERN network for different undersampling mode (e.g., random and radial) and different acceleration factors. Parameter initialization in the LANTERN network：$\rho = 0.2$，$l_r = 0.3$，$\mu_2 = \rho * l_r = 0.06$，$\mu_1 = 1 - \mu_2 = 0.94$，$\tilde{\eta} = 1.8$, stage=13, substage=1, batchsize=1, learning rate of 0.01 and epoch=400.

## Dataset

We collected 101 fully sampled dynamic cardiac MR data using a 3T MRI system (SIEMENS MAGNETOM Trio) with a T1-weighted cine flash sequence. The scan parameters were TR/TE=2.58/49.14ms, number of slices=25, slice thickness=8mm, FOV=280mm, spatial resolution=1.5mm, and sampling matrix size=192×100. We cut all 3D cardiac data into 126×126×16 volumes and perform Fourier transform to obtain the K-space data. By applying corresponding masks, images with different under-sampling patterns were obtained (See Table 1 for details). Finally, 150 cardiac data were generated with 100 data for network training, and 50 for network testing.

## Implementation

Considering that D5C5 is a method based on big data sets, if we use 100 data to train, the reconstruction results are not good enough, and the experimental results also verify the conjecture. So, for the D5C5 method, we cut our data to 126*126*16 and get 3200 data, 2900 data for training, 300 data for testing, and then calculated the average NMSE, PSNR, SSIM. The size of the applied filter is 3*3*9, where the first eight (3*3*8) are DCT and the last one (3*3*1) is TV. The online model training took 45 hours on an Intel Xeon (R) CPU E5-2640 V4 @2.40GHz × 40, 64G.

Table 1 shows the experimental setup in this paper, with a maximum acceleration factor of 11x in 1D Random undersampling mode and a maximum acceleration factor of 15x in 2D Radial undersampling mode. It is worth noting that the acceleration factors in the table are net acceleration factors.

TABLE 1. Different masks and acceleration factors setup

| 1D Random | | | | | | | 2D Radial | | | | | | | |
|---|---|---|---|---|---|---|---|---|---|---|---|---|---|---|
| 2X | 3X | 4X | 5X | 7X | 9X | 11X | 2X | 3X | 4X | 5X | 7X | 9X | 11X | 15X |

# Results

## Impact of dataset size

Figure 2 and Figure 3 show the comparison of the proposed method when the number of training images were increased from 50 to 120. And the average reconstruction quantitative metrics of the 50 test data as shown in Table 2. The proposed method does not require a large amount of data, and can reconstruct a good quality image even in the case of fewer data samples. And from the results, it can be found that the reconstruction result is the best when the data amount is 100. Therefore, we selected 100 images for experiments in subsequent works.

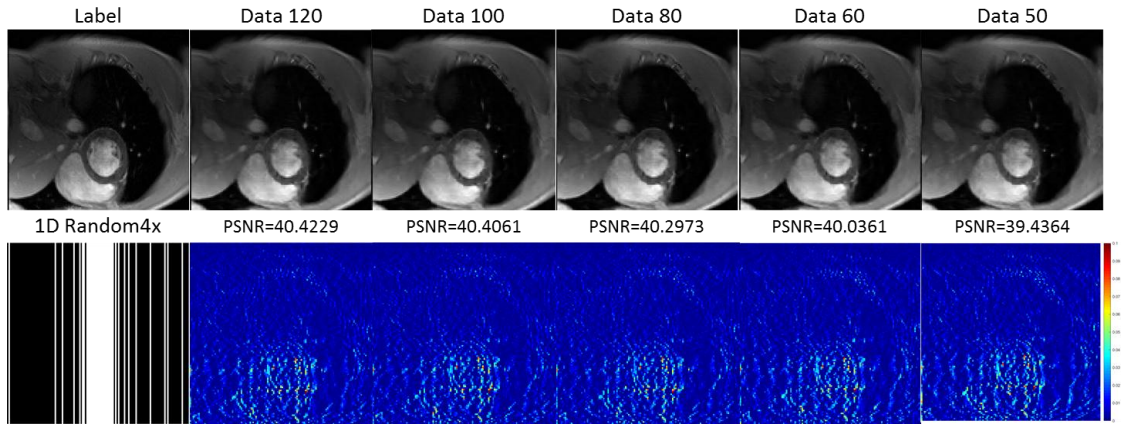

Fig. 2. The comparison of the different amount of data based on the proposed method with 1Drandom sampling at an acceleration factor of 4. PSNR value is given under the results.

TABLE 2. Average reconstruction quantitative metrics of the 50 test data based on different amount of data.

| 1DRandom4x | NMSE | PSNR/dB | SSIM | HFEN |
|---|---|---|---|---|
| data 50 | 0.0413 | 40.8047 | 0.8943 | 0.833 |
| data 60 | 0.0397 | 41.1515 | 0.9 | 0.7939 |
| data 80 | 0.0388 | 41.3589 | 0.9034 | 0.7729 |
| data 100 | **0.0385** | **41.4391** | **0.9043** | **0.7633** |
| data 120 | 0.0386 | 41.4402 | 0.9035 | 0.7685 |

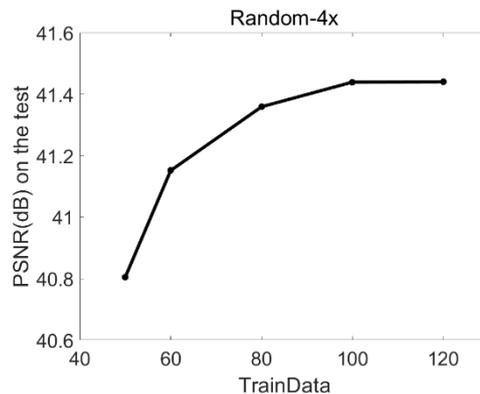

Fig. 3. The comparison between average PSNR value of the 50 test data and different amount of data based on 1D Random sampling at an accelerated factor of 4.

## Effect of initializations: Random Gauss, DCT, DCT+TV

We used the methods of Random Gauss, DCT, and DCT+TV to initialize the proposed model and compare the experiments. Figure 4 shows the reconstruction visual results under several different initialization methods. It can be seen that the reconstruction result initialized by Gaussian noise is the worst, and the reconstruction result under DCT+TV initialization is slightly better than DCT. The PSNR value of DCT+TV is also the highest. Table 3 is the average quantization index corresponding to the reconstruction result under the data volume of 100 and 1Drandom 4x acceleration, which also proves that the DCT+TV method is superior to the other two initialization methods.

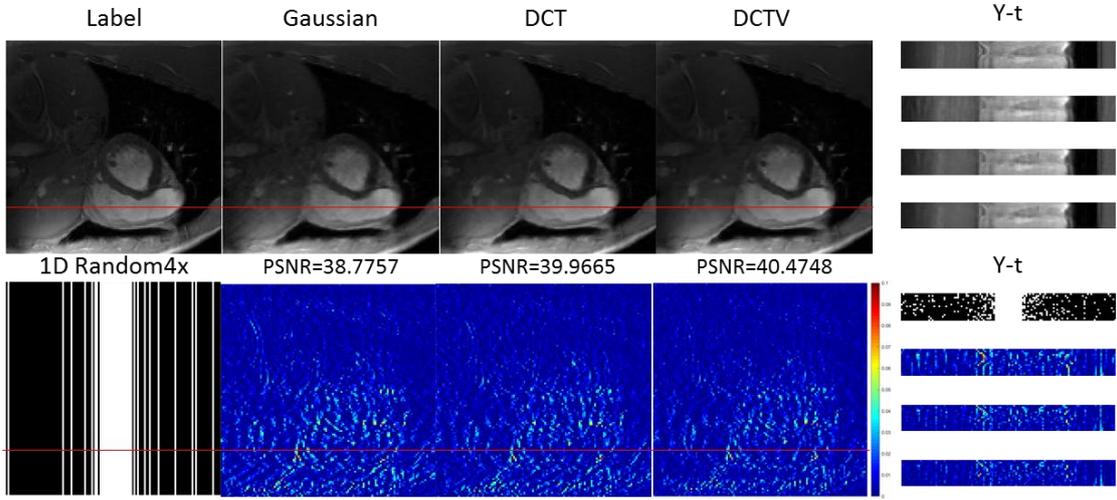

Fig. 4. The comparison of the three initialization modes of Random Gaussian, DCT and LANTERN based on the proposed method with 1Drandom sampling at an acceleration factor of 4. PSNR value is given under the results.

TABLE 3. Average reconstruction quantitative metrics of the 50 test data based on different initiation methods.

| Random4x | Random Gaussian | | | | DCT | | | | DCT+TV | | | |
|---|---|---|---|---|---|---|---|---|---|---|---|---|
| | NMSE | PSNR/dB | SSIM | HFEN | NMSE | PSNR/dB | SSIM | HFEN | NMSE | PSNR/dB | SSIM | HFEN |
| AVE | 0.0461 | 39.8089 | 0.8795 | 0.9459 | 0.0406 | 40.9971 | 0.8946 | 0.8064 | **0.0385** | **41.4391** | **0.9043** | **0.7633** |

## Comparison with state-of-the-art methods

To further verify the feasibility of the proposed model based dynamic MR image reconstruction algorithm, we compare the performance of the proposed method with compressed sensing based k-t SLR technique and data-driven based D5C5 algorithm. It can be seen from the results of 1D random 4x and 5x acceleration factors in Figure 5 and Figure 6 that the reconstructed image by the k-t SLR method is somewhat blurred. The reconstruction visual effect of the D5C5 algorithm is close to the proposed method, but the PSNR value is the highest from the proposed method. Table 4 compares the quantitative indicators of 4x, 7x, and 11x acceleration factors under 1Drandom sampling. We can see that the proposed method has the best index. We also plotted a quantitative index plot of the reconstruction results from 2x to 11x acceleration factors, as shown in Figure 7. The performance of all methods is declining as the acceleration factor increases. However, our approach has always maintained optimal performance.

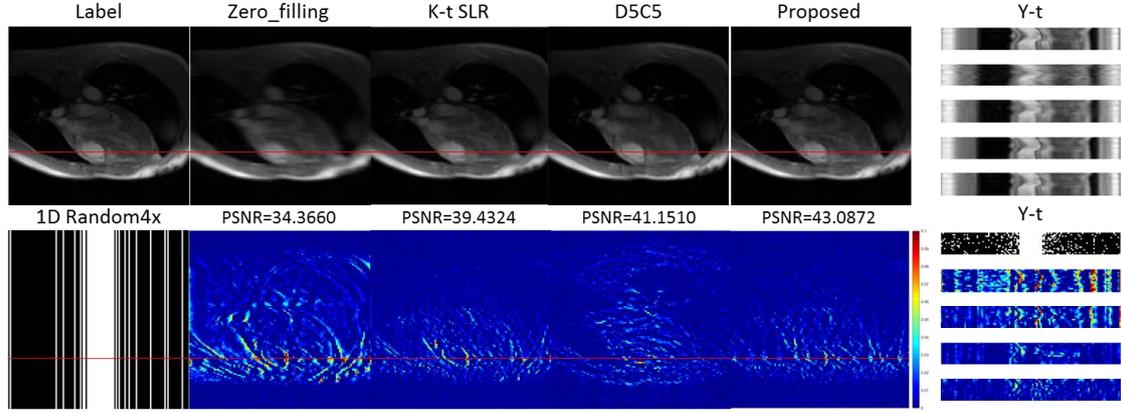

Fig. 5. The comparison of k-t SLR, D5C5 and the proposed method with 1Drandom sampling at an acceleration factor of 4. PSNR value is given under the results.

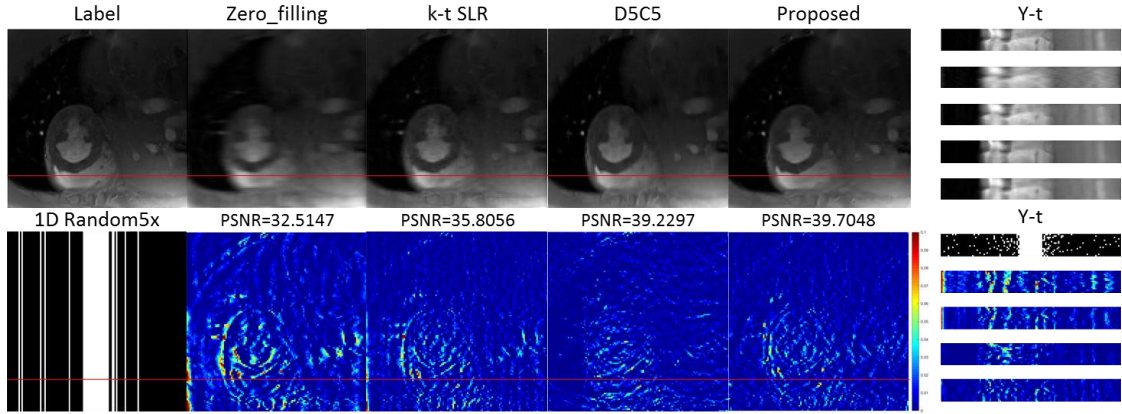

Fig. 6. The comparison of k-t SLR, D5C5 and the proposed method with 1Drandom sampling at an acceleration factor of 5. PSNR value is given under the results.

TABLE 4. Average reconstruction quantitative metrics with standard deviation of the 50 test data based on various methods with 1D Random sampling at a different accelerated factor.

| Methods | Random 4X | | | Random 7X | | | Random 11X | | |
|---|---|---|---|---|---|---|---|---|---|
| | PSNR/dB | SSIM | HFEN | PSNR/dB | SSIM | HFEN | PSNR/dB | SSIM | HFEN |
| Zero_filling | 32.140±2.36 | 0.702±0.03 | 2.045±0.53 | 29.145±2.25 | 0.574±0.04 | 2.735±0.65 | 27.581±2.08 | 0.511±0.04 | 3.121±0.74 |
| Kt-SLR | 36.744±2.77 | 0.855±0.02 | 1.262±0.38 | 33.498±2.70 | 0.775±0.03 | 1.829±0.53 | 32.44±2.61 | 0.733±0.03 | 2.014±0.63 |
| D5C5 | 39.938±2.13 | 0.863±0.03 | 0.831±0.21 | 36.762±2.00 | 0.784±0.03 | 1.401±0.33 | 35.218±2.00 | 0.735±0.03 | 1.817±0.50 |
| Proposed | **41.439**±2.51 | **0.904**±0.02 | **0.763**±0.23 | **37.477**±2.45 | **0.825**±0.02 | **1.309**±0.36 | **35.397**±2.60 | **0.770**±0.03 | **1.668**±0.53 |

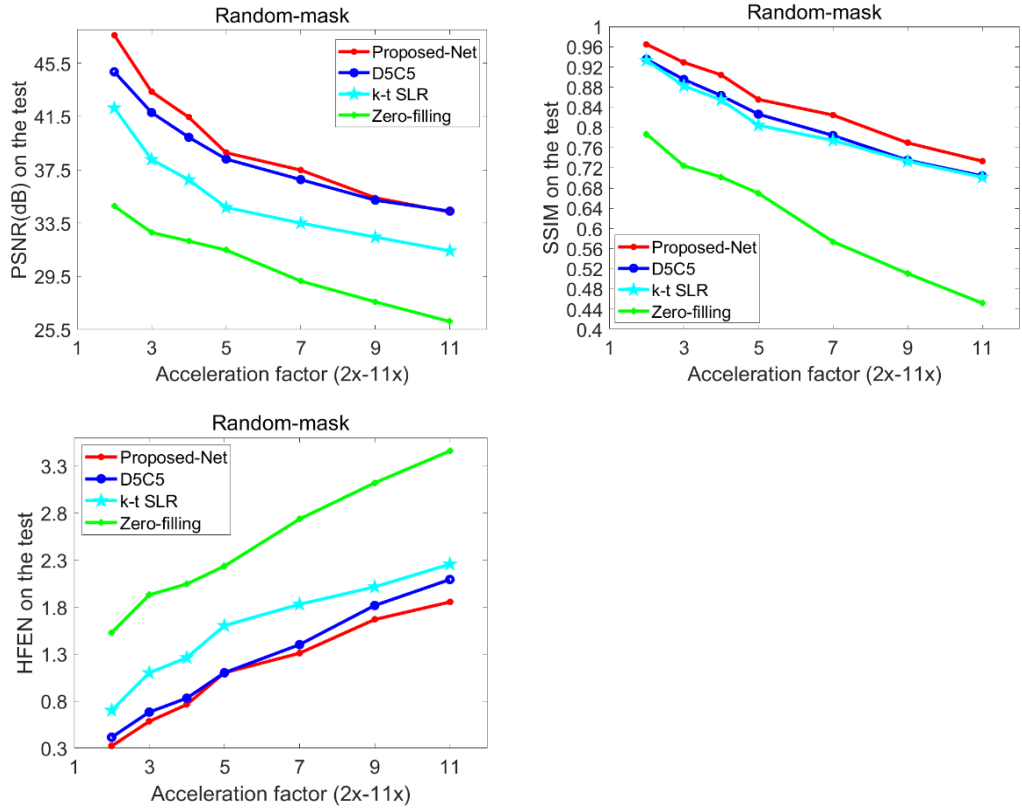

Fig. 7. The comparison of various methods between average quantification index of the 50 test data and acceleration factor based on 1D Random sampling.

In addition, we also conducted an experimental comparison of the radial undersampling trajectory. Table 5 shows the average quantified index values of the test data with the radial sampling at 7x, 11x, and 15x acceleration factors. It shows that our method is far superior to the comparison algorithm in PSNR, SSIM or HFEN. The comparison algorithm D5C5 and k-t SLR have their own advantages. The PSNR and HFEN values of D5C5 are better than k-t SLR, and the SSIM value of k-t SLR is relatively better. We give a visual comparison of the reconstruction results of the radial sampling under the 11x acceleration factor. The reconstruction results of the k-t SLR and D5C5 algorithms are both fuzzy. The proposed method is closest to the original image, and can also be seen from the error map. Fig. 9 is a graph showing the variation of the average quantization index of the test data of several methods with the increase of the acceleration factor, and the results are consistent with those described in Table 5.

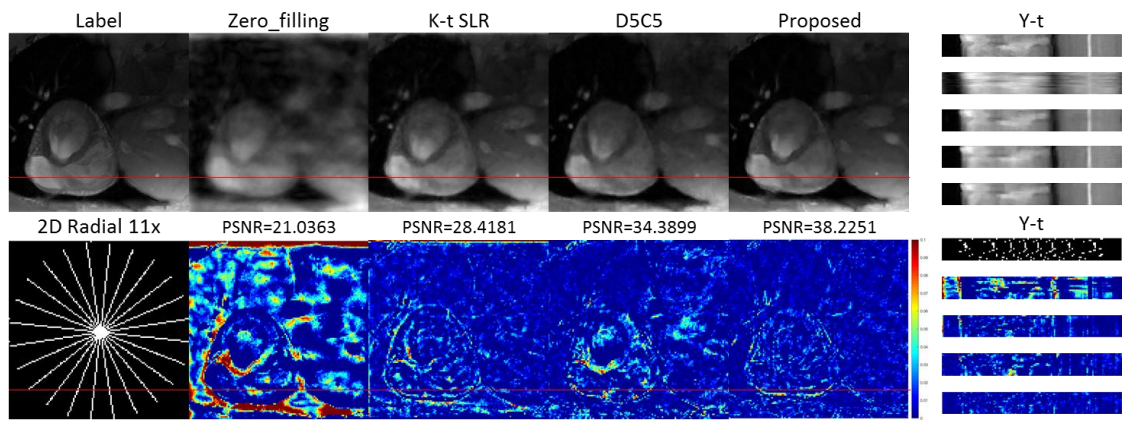

Fig. 8. The comparison of k-t SLR, D5C5 and the proposed method with 2DRadial sampling at an acceleration factor of 11. PSNR value is given under the results.

TABLE 5. Average reconstruction quantitative metrics with standard deviation of the 50 test data based on various methods with Radial sampling at a different accelerated factor.

| Methods | Radial 7X | | | Radial 11X | | | Radial 15X | | |
|---|---|---|---|---|---|---|---|---|---|
| | PSNR/dB | SSIM | HFEN | PSNR/dB | SSIM | HFEN | PSNR/dB | SSIM | HFEN |
| Zero_filling | 26.140±1.45 | 0.467±0.06 | 3.872±0.58 | 22.269±1.37 | 0.345±0.06 | 5.198±0.72 | 20.153±1.27 | 0.275±0.05 | 5.986±0.67 |
| Kt-SLR | 36.032±2.73 | 0.807±0.03 | 1.353±0.37 | 31.961±2.34 | 0.718±0.03 | 2.179±0.51 | 31.518±2.36 | 0.707±0.04 | 2.229±0.54 |
| D5C5 | 38.086±2.18 | 0.800±0.03 | 1.115±0.29 | 34.954±2.08 | 0.701±0.03 | 1.735±0.42 | 34.248±2.04 | 0.677±0.03 | 1.907±0.45 |
| Proposed | **41.440**±2.45 | **0.882**±0.02 | **0.692**±0.20 | **38.874**±2.28 | **0.831**±0.03 | **1.019**±0.26 | **38.115**±2.23 | **0.808**±0.03 | **1.164**±0.30 |

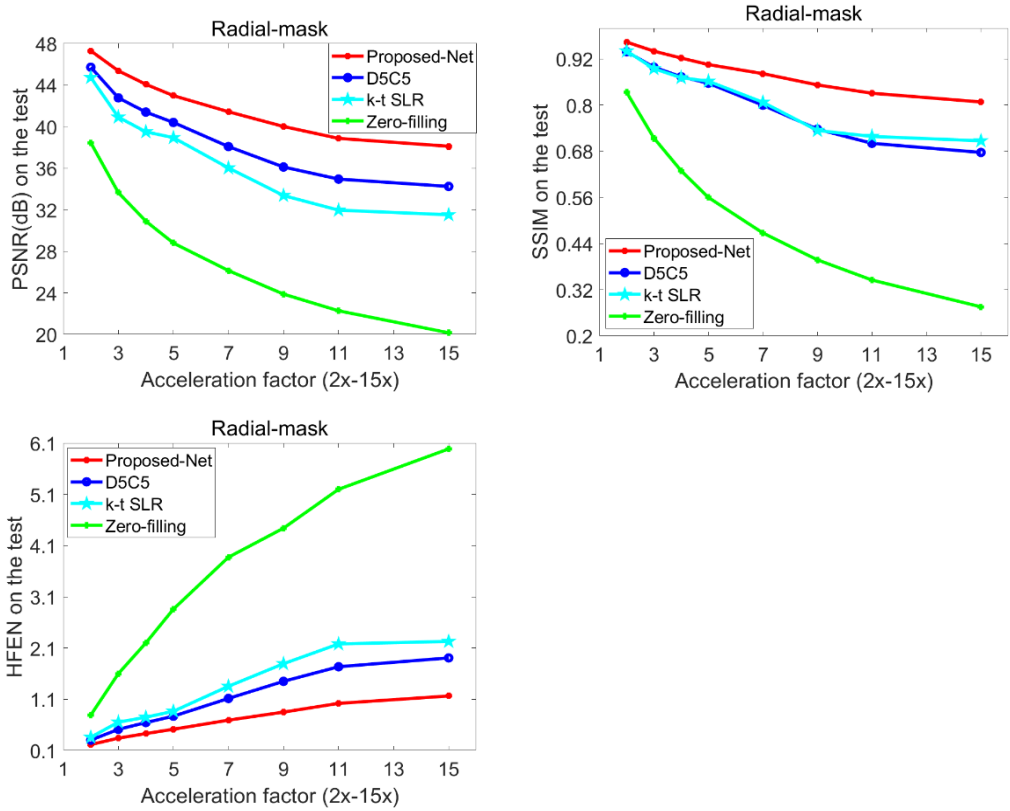

Fig. 9. The comparison of various methods between average quantification index of the 50 test data and acceleration factor based on Radial sampling.

## Convergence analysis

The results in the previous section demonstrate that our method has the best test performance. Due to the small number of samples used, over-fitting problems may occur in machine learning studies. To prove that our method is convergent, there is no over-fitting. We give the training and validation error curves of the proposed model. As shown in Figure 10, a total of 400 epochs are trained, and the training and validation errors are always decreasing, which proves that the proposed method has no over-fitting.

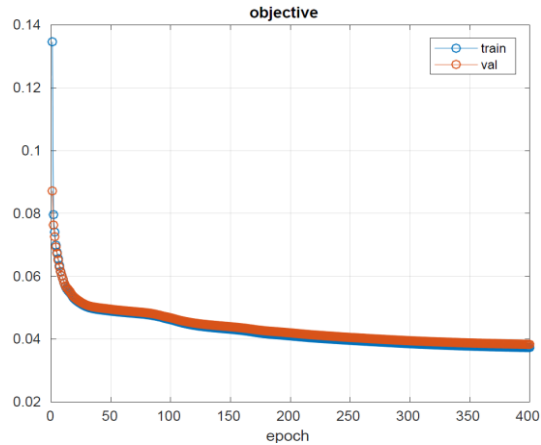

Fig. 10 The training and validation loss curves of the proposed model

# Discussion

Our method simultaneously sparsely constrains the spatial and temporal domains of dynamic magnetic resonance images and uses the end-to-end framework to learn the parameters. By comparing with the most advanced dynamic magnetic resonance image reconstruction algorithm, as shown in Figures 4, 5 and 7, it can be seen from the visual effect that the proposed method can reconstruct clearer results and the image detail content recovery is more accurate. Under the 4x acceleration factor of 1Drandom sampling, the contrast algorithm D5C5 is slightly better than the k-t SLR, and under the 1Drandom sampling 5x acceleration factor, D5C5 is significantly better than the k-t SLR. The reconstruction results of the two contrast algorithms under the 11x acceleration factor of radial sampling have their own advantages and disadvantages. Although D5C5 is higher in PSNR value, D5C5 is more serious for image smoothing, while k-t SLR contains some noise-like artifacts. Comparatively, the reconstruction accuracy of the proposed method is higher. Figures 6 and 8 further demonstrate that the results of the proposed method are superior to the comparison algorithm in terms of various quantitative indicators.

Therefore, due to sparse constraints in space and time, and combining traditional constraint methods with deep learning, the proposed algorithm improves the image reconstruction ability and compensates for the limitations of pure deep learning algorithms that require big data. We can train a better reconstruction model with less sample data than D5C5 algorithm which requires a large amount of data.

In addition, in the reconstruction time, the proposed method reconstructs an image for less than 3s, the reconstruction time of the k-t SLR is about 200s, and the D5C5 is within one second. In terms of time, although it is a little slower than the D5C5, it is still very fast.

# Conclusion

This work proposed a model-based convolutional dynamic MR imaging framework. The framework is able to fully exploit the redundancy in spatial and temporal of dynamic MR images. The experiment results have shown that the proposed method can reconstruct better MR images than the state-of-the-art algorithms in a shorter time under the same acceleration factors.